\documentclass{article}
\usepackage{dcase2023,amsmath,url,times,booktabs,tabularx,amsmath,amssymb}
\usepackage{array,multirow}
\usepackage{arydshln}
\usepackage{multicol}
\usepackage{footnote}
\usepackage{float}
\usepackage{cite}
\usepackage{enumitem}
\usepackage{color,soul}
\usepackage{subcaption}

\usepackage[pdftex]{graphicx}

\newcolumntype{C}[1]{>{\centering\arraybackslash}p{#1}}
\newcolumntype{L}[1]{>{\raggedright\arraybackslash}p{#1}}
\newcolumntype{R}[1]{>{\raggedleft\arraybackslash}p{#1}}

\makeatother
\newcommand{\vect}[1]{{\mbox{\boldmath $#1$}}}
\newcommand{\T}[0]{\mathsf{T}}

\title{DESCRIPTION AND DISCUSSION ON DCASE 2026 CHALLENGE TASK 2: NOISE-AWARE UNSUPERVISED ANOMALOUS SOUND DETECTION FOR MACHINE CONDITION MONITORING}

\name{
Tomoya Nishida$^{1}$, Noboru Harada$^{2}$, Daiki Takeuchi$^{2}$, Daisuke Niizumi$^{3}$, Keisuke Imoto$^{4}$
}
\secondlinename{
Kota Dohi$^{1}$, Harsh Purohit$^{1}$, Takashi Endo$^{1}$, and Yohei Kawaguchi$^{1}$
}

\address{
$^1$ Hitachi, Ltd., Japan, \url{tomoya.nishida.ax@hitachi.com}\\
$^2$ NTT, Inc., Japan, \url{harada.noboru@ntt.com}\\
$^3$ SB Intuitions Corp., \url{nizumical@gmail.com}\\
$^4$ Kyoto University, Japan, \url{keisuke.imoto@ieee.org}\\
}

\begin{document}

\ninept
\maketitle

\begin{sloppy}

\begin{abstract}

This paper presents an overview of DCASE 2026 Challenge Task 2, titled “Noise-aware unsupervised anomalous sound detection (UASD) for machine condition monitoring.”
The task aims to advance noise-robust anomalous sound detection for machine condition monitoring under the unsupervised setting, where only normal machine sounds are available for training.
Reliable detection under noisy conditions is crucial for practical deployment, but previous DCASE Task 2 settings provided limited information about environmental noise, potentially limiting UASD performance in highly noisy situations.
To address this limitation, DCASE 2026 allows participants to exploit two-channel audio samples simultaneously captured at locations near and far from the target machine.
Since the distant microphone is expected to contain relatively stronger environmental noise and weaker direct machine sounds, it may help distinguish environmental noise components from the target machine sounds.
After the challenge submission deadline, challenge results and an analysis of the submitted systems will be added.
\end{abstract}

\begin{keywords}
anomaly detection, acoustic condition monitoring, domain shift, first-shot problem, DCASE Challenge
\end{keywords}

\section{Introduction}
\label{sec:intro}
\vspace{-6pt}
Anomalous sound detection (ASD)~\cite{koizumi2017neyman, kawaguchi2017how, koizumi2019neyman, kawaguchi2019anomaly, koizumi2019batch, suefusa2020anomalous, purohit2020deep} involves determining whether the sound emitted from a target machine is normal or anomalous.
This capability plays a crucial role in automating the detection of mechanical failures, which is essential in the era of the fourth industrial revolution and AI-driven factory automation.

One of the key challenges in developing ASD systems lies in the scarcity and limited diversity of anomalous samples available for training.
To address this, the first ASD task was introduced in the DCASE Challenge 2020 Task 2~\cite{Koizumi2020dcase}, focusing on “unsupervised ASD~(UASD),” which aimed to detect unknown anomalous sounds using only normal sound samples for training. 
Building on this, subsequent challenges in 2021 and 2022~\cite{Kawaguchi2021, Dohi2022DCASE} tackled the issue of domain shifts to enable broader application of ASD systems. 
Domain shifts refer to discrepancies between data from the source and target domains, arising due to variations in machine operational conditions or environmental noise.
The following tasks from 2023 to 2025 (“first-shot” UASD)~\cite{Dohi2023DCASE, Nishida2024DCASE} targeted a realistic setting where systems must detect anomalies for entirely novel machine types without access to similar-type data for training or hyperparameter tuning.
This reflects rapid-deployment scenarios in which collecting diverse training or test data, especially anomalous samples, is infeasible, and therefore manual test-driven tuning is unrealistic.

While various methods have been developed to improve UASD performance under such task settings, achieving high detection performance under noisy conditions remains an important challenge in real-world applications.
To address this, the DCASE 2026 Challenge Task 2 focuses on noise-aware UASD, where two-channel audio samples captured near and far from the target machine are provided.
The distant microphone can serve as a noise-reference signal since it generally contains less direct target machine sounds and relatively stronger environmental noise, which can potentially be leveraged to improve robustness against noise.

After the challenge submission deadline, we will provide results and analysis of the submissions.

\vspace{-6pt}
\section{Noise-aware Unsupervised Anomalous Sound Detection under Domain Shifted Conditions} 
\label{sec:uasd}
\vspace{-6pt}
Consider an audio clip $\vect{x}$, which contains sounds produced by a machine.
The objective of the ASD task is to classify the machine as either normal or anomalous by calculating an anomaly score $\mathcal{A}_{\theta}(\vect{x})$ using an anomaly score calculator $\mathcal{A}$ with parameters $\theta$. 
The input of $\mathcal{A}$ can be the audio clip $\vect{x}$ with or without additional information such as labels indicating the operation condition of the machine.
The machine is then determined to be anomalous when $\mathcal{A}_{\theta}(\vect{x})$ exceeds a pre-defined threshold $\phi$ as
\begin{equation}
\mbox{Decision} = \left\{
\begin{array}{ll}
\mbox{Anomaly} & (\mathcal{A}_{\theta}(\vect{x}) > \phi)\\
\mbox{Normal} & (\mbox{otherwise}).
\end{array}
\right.
\label{eq:det}
\end{equation}
The primary difficulty in this task is to train the anomaly score calculator with only normal sounds (UASD). 
The DCASE 2020 Challenge Task 2~\cite{Koizumi2020dcase} was designed to address this issue, and all the following tasks stand on this UASD setting.
In addition to this basic UASD setting, this year's challenge has three key features essential for the practical implementation of ASD systems.
The first two features were introduced in the previous challenges, and the third feature is newly introduced in this year's challenge.

The first feature is addressing the domain-shift problem.
Domain shifts refer to variations in conditions between training and testing phases, which alter the distribution of the observed sound data.
These variations can result from differences in operating speed, machine load, heating temperature, microphone arrangement, environmental noise, and other factors.
Two domains are defined: the \textbf{source domain}, representing the original condition with sufficient training data, and the \textbf{target domain}, representing another condition where only limited samples are available.
This year's task follows the 2022 to 2025 Task 2~\cite{Dohi2022DCASE, Dohi2023DCASE, Nishida2024DCASE} setting, where the domain information is assumed to be unknown in the test phase and anomalies from both domains have to be detected with a single threshold.
In this case, domain generalization is required to achieve good performance.

The second feature is addressing the "first-shot problem".
For the rapid development of ASD systems in real-world scenarios, solving ASD (a) against completely novel machine types (b) with only one section of training data (c) without handcrafted tunings that depend on test data, is highly important.
This is because in real-world scenarios, customers may only possess a single novel machine, and collecting test data, especially anomalous samples, for handcrafted tuning may be infeasible.
This problem setting is referred to as the “first-shot problem”, and the Task 2 2023 to 2025~\cite{Dohi2023DCASE, Nishida2024DCASE, Nishida2025DCASE} was organized based on this problem setting.
This was implemented by introducing two features to the dataset:
(i) The development dataset (which contains both training and test data with ground truth labels for the participants to develop their systems) and evaluation datasets (which do not contain ground truth labels, and are used for final evaluation of the systems) consist of entirely different sets of machine types, and
(ii) Each machine type in the dataset contains only a single section.

The third feature is the focus on noise robustness.
Detecting anomalous machine sounds under high noise conditions is an important problem, since real-world environments often contain various sources of noise.
While the previous DCASE Task 2 Challenges have included noisy conditions, achieving high detection performance under such conditions remained challenging in some cases.
To address this issue and encourage the development of noise-robust UASD systems, this year's task provides additional information related to noise.
Specifically, whereas previous DCASE Challenge Task 2 datasets provided single-channel audio, this year's training and test samples are provided as two-channel recordings simultaneously captured by microphones placed at locations near to and far from the target machine. 
Participants may use either or both channels as input to their system.
Since the distant microphone is expected to contain less sound from the target machine and relatively stronger environmental noise, it may provide useful cues for identifying which components of the recording correspond to environmental noise.
Note that this setting is complementary to the DCASE 2025 setting.
In DCASE 2025, supplemental data for improving noise robustness were provided in the form of either clean recordings of the target machine sounds or recordings containing only noise.
Such a setting is applicable when the target machine or the noise sources can be stopped so that such recordings can be obtained in advance.
However, in some real-world scenarios, such recordings may be difficult to obtain.
In contrast, the DCASE 2026 setting is applicable where such recordings are unavailable but two microphones can be installed.

\section{Task Setup} 
\label{sec:task}
\vspace{-6pt}

\subsection{Dataset} 
\label{sec:dataset}
\vspace{-5pt}
The dataset for this task is divided into three categories: the \textbf{development dataset}, the \textbf{additional training dataset}, and the \textbf{evaluation dataset}. 
The development dataset contains seven machine types, while the additional training and evaluation datasets include five machine types, with each machine type consisting of a single section.
A \textbf{machine type} refers to the category of machines, such as fans or gearboxes, and a \textbf{section} represents a subset or the entirety of the data associated with each machine type.

All recordings are two-channel, lasting 6 to 16 seconds, and have a sampling rate of 16 kHz. 
The machine sounds were recorded in a laboratory environment using two microphones placed at different distances from the target machine, with one microphone positioned close to the machine and the other farther away.
Environmental noise was first recorded in factories or suburban areas, then played back through loudspeakers in the laboratory and rerecorded using the same microphone setup as that used for the machine-sound recordings.
The microphone setup, such as the distance between the microphones and the target machine, and laboratory arrangement differ across machine types, but the same setup was used consistently within each machine type.

The \textbf{development dataset} provides seven machine types (fan, gearbox~(Emu), bearing~(Emu), slide rail~(Emu), valve~(Emu), ToyCar, ToyCar~(Emu)), and each machine type has one section that contains a complete set of the training and test data.
Each section contains
(i) 990 normal clips from a source domain for training, 
(ii) 10 normal clips from a target domain for training, and
(iii) 100 normal clips and 100 anomalous clips from both domains for the test.
To assist participants, domain information (source/target) was included in the test data. 
For three machine types (ToyCar~(Emu), fan, gearbox) details regarding the operational or environmental conditions were provided in the file names and attribute CSV files. 
For the remaining four machine types, these attributes were not disclosed.
For five machine types (gearbox~(Emu), bearing~(Emu), slide rail~(Emu), valve~(Emu), ToyCar~(Emu)), impulse responses were measured from the target machine position and from each loudspeaker position to the two microphones. 
The audio was then simulated by convolving these impulse responses with previously recorded machine sounds and environmental noise.
All other data were recorded in real conditions using the setup described above.

The \textbf{additional training dataset} provides novel five machine types (ToyDrone, ToothBrush, SewingMachine, Sander, BlowerDustCollector).
Each section consists of 
(i) 990 normal clips in a source domain for training, and
(ii) 10 normal clips in a target domain for training.
For three machine types (ToyDrone, Sander, BlowerDustCollector), attributes were provided in this dataset.
For the other two machine types (ToothBrush, SewingMachine), attributes were concealed.
For all machine types in this dataset, the audio clips were directly recorded in the laboratory as described above.
The \textbf{evaluation dataset} provides the test clips that correspond to the additional training dataset, e.g. data of the same machine types as the additional training dataset. 
Each section consists of 200 test clips, none of which have ground-truth labels (i.e., normal or anomaly), domain information, or attribute information. 
Participants are required to train a model for each new machine type using only a single section per machine type.

\subsection{Evaluation metrics} 
\label{sec:metrics}
\vspace{-5pt}
To assess overall detection performance, we employed the area under the receiver operating characteristic curve (AUC). 
Additionally, we used the partial AUC~(pAUC) to evaluate performance in a low false-positive rate range $[0, p]$, where we set $p = 0.1$.
Let $m$ and $n$ denote the index of a machine type and a section respectively, and $d \in \{ {\rm source}, {\rm target} \}$ a domain.
To evaluate each system under the domain generalization setting, we compute the AUC for each domain and pAUC for each section as
\begin{equation}
	{\rm AUC}_{m, n, d} = \frac{1}{N^{-}_{d}N^{+}_{n}} \sum_{i=1}^{N^{-}_{d}} \sum_{j=1}^{N^{+}_{n}}
	\mathcal{H} (\mathcal{A}_{\theta} (x_{j}^{+}) - \mathcal{A}_{\theta} (x_{i}^{-})),
\end{equation}
\begin{equation}\text{\scalebox{0.93}{$
	{\rm pAUC}_{m, n} = \frac{1}{\lfloor p N^{-}_{n} \rfloor N^{+}_{n}} \sum_{i=1}^{\lfloor p N^{-}_{n} \rfloor N^{+}_{n}} \sum_{j=1}^{N^{+}_{n}}
	\mathcal{H} (\mathcal{A}_{\theta} (x_{j}^{+}) - \mathcal{A}_{\theta} (x_{i}^{-})),
$}}
\end{equation}
where $\lfloor \cdot \rfloor$ is the flooring function and $\mathcal{H} (y)$ returns 1 when $y > 0$ and 0 otherwise.
Here, $\{x^{-}_{i}\}_{i=1}^{N^{-}_{d}}$ are the normal test clips in domain $d$ in section $n$ of machine type $m$ and $\{x_{j}^{+}\}_{j=1}^{N^{+}_{n}}$ are all the anomalous test clips in section $n$ of machine type $m$.
$N^{-}_{d}, N^{-}_{n}, N^{+}_{n}$ represent the number of normal test clips in domain $d$, normal test clips in section $n$, and anomalous test clips in section $n$, respectively.

The official score $\Omega$ is given by the harmonic mean of the AUC and pAUC scores overall machine types and sections:
\begin{eqnarray}
\Omega &=& h \left\{ {\rm AUC}_{m, n, d}, \ {\rm pAUC}_{m, n} \quad | \quad \right. \nonumber \\
&& \left. m \in \mathcal{M}, \  n \in \mathcal{S}(m), \ d \in \{ {\rm source}, {\rm target} \} \right\},
\end{eqnarray}
where $h\left\{\cdot\right\}$ represents the harmonic mean, $\mathcal{M}$ is the set of given machine types, and $\mathcal{S}(m)$ represents the set of sections for machine type $m$.
Since the dataset in 2024, $\mathcal{S}(m)$ contains only $00$.

\subsection{Baseline systems and results}
\label{sec:baseline}
\vspace{-5pt}
The task organizers offer a baseline system using Autoencoders (AEs) with two operating modes, identical to the 2023 Task 2 baseline.
While this year's data contains two-channel recordings, the baseline system uses only the first channel as input and does not utilize the second channel.
While both modes use Autoencoders for training, they differ in anomaly score computation.
This paper presents the system and its detection performance; details can be found in \cite{Harada2023}.

\subsubsection{Autoencoder training}
\vspace{-5pt}
The AE is trained for both operating modes using log-mel-spectrograms of training sound clips $X = [X_1, \dots, X_T]$, where $X_t \in \mathbb{R}^F$ for $t=1,\dots,T$ represents frame-wise feature vectors at frame $t$, where $F=128$ and $T$ is the number of mel-filters and time-frames, respectively.
For input, $P=5$ consecutive frames are concatenated as $\psi_t = [X_t^\T, \dots, X_{t + P - 1}^\T]^\T \in \mathbb{R}^{D}$ for each $t$, with $D=P \times F = 640$.
Model parameters are trained by minimizing the mean squared error~(MSE) between the input $\psi_t$ and the reconstructed output $r_\theta (\psi_t)$ for all inputs from the training data.

\subsubsection{Simple Autoencoder mode}
\vspace{-5pt}
This mode uses the mean MSE of all features derived from the given sound clip as its anomaly score, e.g.,
\begin{equation}
A_{\theta}(X) = \frac{1}{DK} \sum_{k = 1}^K \| \psi_k - r_{\theta}(\psi_k) \|_{2}^{2},
\end{equation}
where $K=T-P+1$, and $\| \cdot \|_2$ represents $\ell_2$ norm.

\subsubsection{Selective Mahalanobis mode}
\vspace{-5pt}
In this mode, the Mahalanobis distance between the system input and reconstructed feature is used to compute the anomaly score. 
The anomaly score is defined as
\begin{align}
&\text{\scalebox{0.95}{$A_{\theta}(X) = \frac{1}{DK} \sum_{k = 1}^K \min\{ D_s (\psi_k, r_{\theta}(\psi_k)), D_t (\psi_k, r_{\theta}(\psi_k))\}$}},\\
&D_s(\cdot) = \textrm{Mahalanobis}(\psi_k, r_{\theta}(\psi_k), \Sigma_s^{-1}), \\
&D_t(\cdot) = \textrm{Mahalanobis}(\psi_k, r_{\theta}(\psi_k), \Sigma_t^{-1}),
\end{align}
where $\Sigma_s^{-1}$ and $\Sigma_t^{-1}$ are the covariance matrices of $r_{\theta}(\psi_k) - \psi_k$ for the source and target domain data of each machine type, respectively.

\subsubsection{Results}
\label{sec:results}
\setlength{\tabcolsep}{1mm}

\begin{table}[t]
\begin{center}
\vspace{-10pt}
\caption{Baseline results for development dataset.}
\label{tab:baseline_results}
\scriptsize
\begin{tabular}{@{}l c c c p{1pt} c c@{}}
\hline
\ \\[-6.5pt]
Machine type &
Mode &
\multicolumn{2}{c}{AUC [\%]} &&
\multicolumn{2}{c}{pAUC [\%]} \\
\cline{3-4} \cline{6-7}
\ \\[-6.5pt]
& & 
\multicolumn{1}{c}{Source} &
\multicolumn{1}{c}{Target} && \\
\hline
\ \\[-6.5pt]
  ToyCar~(Emu)
  &	MSE & $69.62 \pm 9.95$ & $61.20 \pm 6.34$ && $55.89 \pm 3.64$\\
  
  &	MAHALA & $69.49 \pm 1.71$ & $66.62 \pm 6.74$ && $53.47 \pm 2.28$\\
  \hline
  
  ToyCar
  &	MSE & $75.62 \pm 1.98$ & $37.87 \pm 1.37$ && $54.03 \pm 0.70$\\
  
  &   MAHALA & $77.28 \pm 1.57$ & $53.17 \pm 2.98$ && $58.25 \pm 0.53$\\
  \hline
  
  bearing~(Emu)
  &	MSE & $62.34 \pm 1.09$ & $59.56 \pm 0.83$ && $59.85 \pm 0.23$\\
  
  &	MAHALA & $65.92 \pm 1.74$ & $62.28 \pm 1.31$ && $60.42 \pm 0.37$\\
  \hline
  
  fan
  &	MSE & $61.45 \pm 0.66$ & $46.94 \pm 0.52$ && $53.33 \pm 0.37$\\
  
  &	MAHALA & $60.00 \pm 4.09$ & $45.09 \pm 1.76$ && $52.29 \pm 0.32$\\
  \hline
  
  gearbox~(Emu)
  &	MSE & $68.23 \pm 1.71$ & $49.78 \pm 0.75$ && $52.94 \pm 0.69$\\
  
  &	MAHALA & $74.48 \pm 2.01$ & $52.74 \pm 1.90$ && $53.97 \pm 0.59$\\
  \hline
  
  slider~(Emu)
  &	MSE & $67.25 \pm 1.11$ & $45.05 \pm 0.95$ && $50.38 \pm 0.40$\\
  
  &	MAHALA & $66.36 \pm 0.45$ & $49.18 \pm 0.51$ && $50.36 \pm 0.17$\\
  \hline
  
  valve~(Emu)
  &	MSE & $67.74 \pm 1.45$ & $68.78 \pm 1.03$ && $55.08 \pm 0.80$\\
  
  &	MAHALA & $56.60 \pm 1.39$ & $56.50 \pm 1.49$ && $50.20 \pm 0.70$\\
  \hline
\end{tabular}
\end{center}
\end{table}

Tables \ref{tab:baseline_results} present the AUC and pAUC results for the two baseline systems on the development dataset, with the averages and standard deviations computed from five independent trials.

\section{Challenge Results}
We will provide Challenge results and analysis of the submissions to the DCASE 2026 Workshop.

\section{Conclusion}
We presented an overview of the DCASE 2026 Challenge Task 2.
The task aims to develop ASD systems that work under noisy conditions, given two-channel audio recordings recorded at different distances from the target machine.
Results and analysis of the challenge submissions will be added following the challenge submission deadline.

\clearpage
\bibliographystyle{IEEEtran}
\bibliography{refs}

\end{sloppy}
\end{document}